\documentclass[preprint,aps,nofootinbib]{revtex4}
\usepackage[dvipdfmx]{graphicx}
\usepackage{color,amssymb,amsmath,ascmac}
\usepackage{slashed}

\usepackage{amsmath}
\usepackage{amsfonts}
\usepackage{amssymb}

\usepackage{amsmath}
\usepackage{amsfonts}
\usepackage{amssymb}
\usepackage{color}
\usepackage{hyperref}

\def\be{\begin{equation}}
\def\ee{\end{equation}}
\def\bea{\begin{eqnarray}}
\def\eea{\end{eqnarray}}

\def\ss2l{SS2$\ell$}
\def\3l{3$\ell$}

\begin{document}

\newcommand{\lrf}[2]{ \left(\frac{#1}{#2}\right)}
\newcommand{\lrfp}[3]{ \left(\frac{#1}{#2} \right)^{#3}}
\newcommand{\vev}[1]{\left\langle #1\right\rangle}

\newcommand{\TeV}{\text{TeV}}
\newcommand{\GeV}{\text{GeV}}
\newcommand{\MeV}{\text{MeV}}
\newcommand{\keV}{\text{keV}}
\newcommand{\eV}{\text{eV}}

%%%%%%%%%%%%%%%%%%%%%%%%%%%%%%%%%%%%%%%%
\title{From the 750 GeV Diphoton Resonance to Multilepton Excesses}
\vspace*{1cm}
%%%%%%%%%%%%%%%%%%%%%%%%%%%%%%%%%%%%%%%%

\author{\vspace{1cm} Kyu Jung Bae$^{\, a}$, Chuan-Ren Chen$^{\, b}$, Koichi Hamaguchi$^{\, a,c}$ and   Ian Low$^{\, d,e}$ }

\affiliation{
\vspace*{.5cm}
  \mbox{$^a$Department of Physics, University of Tokyo, Bunkyo-ku, Tokyo 113--0033, Japan}\\
  \mbox{$^b$Department of Physics, National Taiwan Normal University, Taipei 116, Taiwan}\\
  \mbox{$^c$Kavli Institute for the Physics and Mathematics of the Universe (Kavli IPMU),}\\
  \mbox{University of Tokyo, Kashiwa 277--8583, Japan}\\
\mbox{$^d$ Department of Physics and Astronomy, Northwestern University, Evanston, IL 60208, USA} \\
\mbox{$^e$ High Energy Physics Division, Argonne National Laboratory, Argonne, IL 60439, USA}\\
\vspace*{1cm}}

\begin{abstract}
Weakly-coupled models for the 750 GeV diphoton resonance often invoke new particles carrying both  color and/or electric charges to mediate loop-induced couplings of the resonance to two gluons and two photons. The new colored particles may not be stable and could decay  into final states containing standard model particles. We consider an electroweak doublet of vector-like quarks (VLQs) carrying electric charges of 5/3 and 2/3, respectively, which mediate the loop-induced couplings of the 750 GeV resonance. If the  VLQ has a mass at around 1 TeV, it naturally gives rise to the observed diphoton signal strength while all couplings remain perturbative up to a high scale. At the same time, if the  charge-5/3 VLQ decays into final states containing top quark and $W$ boson, it would contribute to the  multilepton excesses observed in both Run 1 and Run 2 data. It is also possible to incorporate a dark matter candidate in the decay final states to explain the observed relic density.

\end{abstract}

%\pacs{}

\maketitle

%%%%%%%%%%%%%%%%%%%%%%%%%%%%%%%%%%%%%%%%
\section{Introduction}
%%%%%%%%%%%%%%%%%%%%%%%%%%%%%%%%%%%%%%%%

After the discovery of the 125 GeV Higgs boson at Run 1 of the Large Hadron Collider (LHC) \cite{Aad:2012tfa,Chatrchyan:2012xdj}, the standard model (SM) of particle physics is now complete. However, SM is far from being perfect, since there is much empirical evidence pointing to  its failures, ranging from the compelling case for cold dark matter to the baryon asymmetry in the universe. The evidence strongly calls for existence of physics beyond the SM (BSM). 

Not surprisingly, there are many measurements exhibiting excesses over SM expectations at LHC Run 1. While some, or perhaps all, of the excesses may be  statistical fluctuations, it is difficult to over-emphasize the importance of pursuing the theoretical implications of the first signs of BSM physics, which if remains true would be a discovery that is more profound and no less grand than that of  the Higgs boson. 

Among the Run 1 excesses, there is one which seems to persist in the early Run 2 results, showing up in searches for final states containing multilepton, $b$-jets, and missing transverse momentum (MET).  The multilepton searches are often further divided into a same-sign dilepton (2L) category, a three-lepton (3L) category, and a four-lepton (4L) category. As was pointed out in Refs.~\cite{Huang:2015fba,Chen:2015jmn}, the multilepton excesses were observed with varying degrees of significance in many  Run 1 analyses in both ATLAS and CMS collaborations, which include searches for ttH production \cite{Aad:2015iha,Khachatryan:2014qaa}, for scalar bottom quarks in supersymmetry \cite{Aad:2014pda,Chatrchyan:2013fea}, for SM production of ttW \cite{Aad:2015eua,Khachatryan:2015sha} and for heavy VLQs \cite{Aad:2015gdg}. (The CMS Run 1 search for VLQ's, which imposes very hard cuts on the kinematics, is the only exception which didn't see an excess \cite{Chatrchyan:2013wfa}.)  At Run 2, the multilepton excesses were seen by ATLAS in searches for scalar bottom quarks \cite{Aad:2016tuk} and SM ttW production \cite{atlasrun2ttw}. CMS, which collected 50\% less ``good" data than the ATLAS at Run 2 so far, sees a deficit in 2L channel and an excess in the 3L channel \cite{CMS:2016rnk,CMS:2015alb}.

\begin{figure}[t]
  \begin{center}
\includegraphics[width=0.47\textwidth]{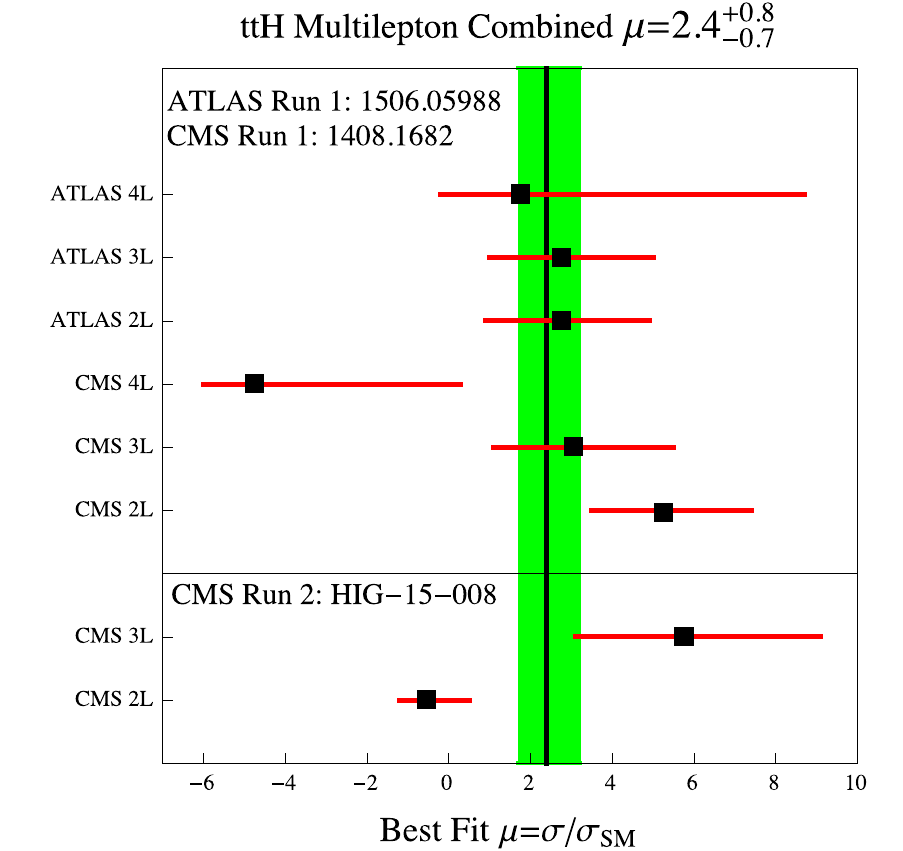}
\includegraphics[width=0.41\textwidth]{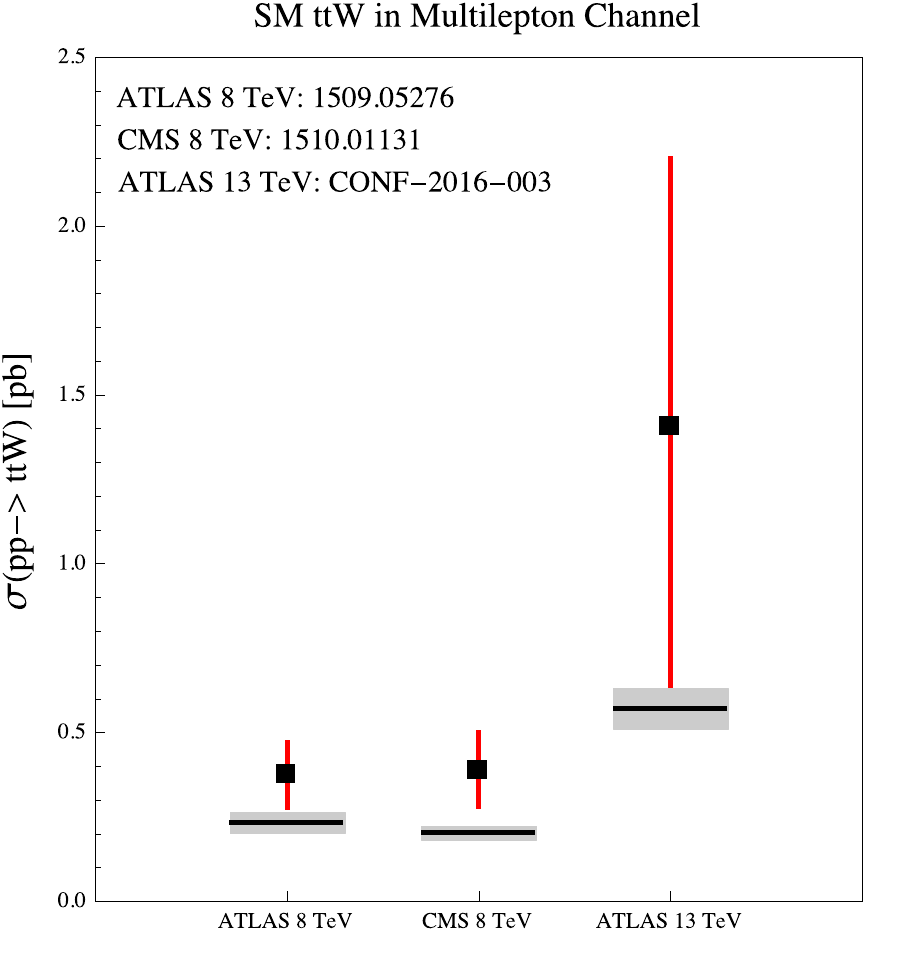}
\caption{\em Left: Best-fit  signal strength $\mu=\sigma/\sigma_{\rm SM}$ in the ttH multilepton channel at both 8 TeV and 13 TeV LHC. Right: Measured $\sigma(pp\to ttW)$ in the multilepton channel at both 8 TeV and 13 TeV LHC. Shaded areas are the expected cross-sections with theoretical uncertainties.
 }{\label{fig:statcomb}}
 \end{center}
\end{figure}

The left panel of Fig.~\ref{fig:statcomb} summarizes  the best-fit signal strength $\mu=\sigma/\sigma_{\rm SM}$ in the ttH multilepton channel from the public Run 1 and Run 2 results, as well as a statistical combination of $\mu$  obtained from using the online script at Ref.~\cite{Barlow:online}, which is based on Ref.~\cite{Barlow:2004wg}. The resulting signal strength is $\mu=2.4_{-0.7}^{+0.8}$ from combining both ATLAS and CMS Run 1/2 results.\footnote{We have added an offset of $-0.3$ in the combined central value to take into account correlated systematic uncertainties. This procedure reproduces the individual central values from ATLAS and CMS, respectively, with good precision.} In the right panel of Fig.~\ref{fig:statcomb} we display measurements of $\sigma(pp\to t\bar{t}W)$ from ATLAS, at both Run 1 and 2, and CMS at Run 1. The presence of excesses in the multilepton channel at both Run 1 and Run 2 is clear, although uncertainties in the measurements remain large. A number of works have theorized on the nature of these excesses \cite{Huang:2015fba,Angelescu:2015kga,Chen:2015jmn,Badziak:2016exn}.

%%%%%%%%%%%%%%%%%%%%%%%%%%%%%%%%%%%%%%%%%%%
%%%%%%%%%%%%%%%%%%%%%%%%%%%%%%%%%%%%%%%%%%%

Nevertheless, the most significant sign of BSM physics in Run 2 undoubtedly comes from the observation of excessive events in the diphoton channel that are clustered at around an invariant mass of 750 GeV \cite{ATLAS:diphoton,CMS:2015dxe}. Moreover, kinematic distributions of events in the signal region seem consistent with those expected from SM background \cite{ATLAS:diphoton2}, indicating single production of a resonance, with no or very soft extra particles, decaying into diphoton final states. On the other hand, compatibility with the null result from ATLAS Run 1 searches for a heavy resonance in the diphoton channel \cite{Aad:2015mna} prefers a production mechanism from the gluon initial states, due to the larger increase in the parton luminosity of the gluon.

If one further assumes the resonance is a spin-0 boson, then in weakly-coupled theories the neutral scalar couplings to massless gauge bosons such as the photon and the gluon are induced only at  one-loop level by particles carrying QCD color and/or electric charges (see, for example, Ref.~\cite{Djouadi:2005gi}.) The particles in the loop,  however, cannot be the SM particles, for it would immediately imply tree-level decays of the resonance into the SM particles, which in turn would swamp the loop-induced decays into diphotons and reduce the signal strength far below the observed ones. Therefore, in these scenarios additional new particles, other than the 750 GeV scalar itself, must be present. These new particles should carry color and/or electric charges and be heavier than 375 GeV so as to turn off the tree-level decay channel of the 750 GeV resonance. A popular possibility for the new particles mediating the loop-induced couplings is VLQ's carrying QCD color and electric charges~\cite{Angelescu:2015uiz,Knapen:2015dap,Buttazzo:2015txu, Pilaftsis:2015ycr,Franceschini:2015kwy,McDermott:2015sck,Ellis:2015oso,Gupta:2015zzs,Molinaro:2015cwg,Dutta:2015wqh,Kobakhidze:2015ldh,Martinez:2015kmn,Chao:2015ttq, Chakrabortty:2015hff,Agrawal:2015dbf,Falkowski:2015swt,Aloni:2015mxa,Benbrik:2015fyz,Chao:2015nsm,Chakraborty:2015jvs,Han:2015dlp,Feng:2015wil,Antipin:2015kgh,Wang:2015kuj,Dhuria:2015ufo,Boucenna:2015pav,Hernandez:2015ywg,Dey:2015bur,Pelaggi:2015knk,Huang:2015rkj,Patel:2015ulo,Badziak:2015zez,Cao:2015xjz,Altmannshofer:2015xfo,Cvetic:2015vit,Gu:2015lxj,Craig:2015lra,Das:2015enc,Cheung:2015cug,Zhang:2015uuo,Hall:2015xds,Han:2015yjk,Salvio:2015jgu,Li:2015jwd,Son:2015vfl,Tang:2015eko,Cao:2015apa,Wang:2015omi,Cai:2015hzc,Cao:2015scs,Goertz:2015nkp,Dev:2015vjd,Bizot:2015qqo,Kang:2015roj,Low:2015qho,Hernandez:2015hrt,Jiang:2015oms,Jung:2015etr,Palti:2016kew,Ko:2016lai,Chao:2016mtn,Karozas:2016hcp,Hernandez:2016rbi,Modak:2016ung,Dutta:2016jqn,Deppisch:2016scs,Bhattacharya:2016lyg,Ko:2016wce,Cao:2016udb,Yu:2016lof,Ding:2016ldt,Faraggi:2016xnm,Han:2016bvl,King:2016wep,Bertuzzo:2016fmv,Salvio:2016hnf,Ge:2016xcq,Godunov:2016kqn,Arbelaez:2016mhg,Gross:2016ioi,Han:2016fli,Staub:2016dxq,Cvetic:2016omj,Ren:2016gyg,Chiang:2016eav,Lazarides:2016ofd,Bonilla:2016sgx,Li:2016tqf,Barbieri:2016cnt,Belanger:2016ywb,Perelstein:2016cxy,Fraser:2016tvn,Huong:2016kpa,Leontaris:2016wsy,DiChiara:2016dez,Csaki:2016kqr,Liu:2016lkj,Gherghetta:2016fhp,Nilles:2016bjl,Duerr:2016eme,Gopalakrishna:2016tku,Chang:2015bzc,Djouadi:2016eyy,Kawamura:2016idj,Hamada:2016vwk,Bae:2016xni,Hamada:2015skp}.

There have been few discussions on the collider phenomenology 
of VLQs associated with the 750 GeV scalar~\cite{Chang:2015bzc,Djouadi:2016eyy,Kawamura:2016idj,Bae:2016xni,Hamada:2016vwk}, 
although VLQs have been proposed in other contexts, as partners of the third generation quarks in the SM \cite{Cheng:2005as,Contino:2008hi}, which could decay into third generation quarks and $W$, $Z$, or the Higgs bosons. In this work we aim to demonstrate that, if the VLQ is an electroweak doublet with hypercharge 7/6, and has a mass at around 1 TeV, it could simultaneously explain the signal strength of the 750 GeV diphoton resonance and contribute to the aforementioned multilepton excesses associated with $b$-jets and MET. And the model could remain perturbative all the way up to a very high scale \cite{Bae:2016xni,Hamada:2015skp}.

This work is organized as follows: in the next section we demonstrate the connection between VLQs and the 750 GeV diphoton resonance, as well as the RG running of the relevant gauge and Yukawa couplings, which is followed by a Monte Carlo study on the contribution of the VLQs to the multilepton excesses at the LHC, using the ttH and ttW channels as examples. In the last section we conclude.

%%%%%%%%%%%%%%%%%%%%%%%%%%%%%%%%%%%%%%
\section{Vector-Like Quarks and Diphoton Excess}
\label{sec:VLQ_and_diphoton}
%%%%%%%%%%%%%%%%%%%%%%%%%%%%%%%%%%%%%%

Assuming that the 750 GeV scalar $S$ is produced by the gluon fusion, the diphoton signal rate is given by 
\begin{align}
\sigma(pp\to S\to \gamma\gamma)
&\simeq
\frac{C_{gg}}{s\cdot m_S}\frac{\Gamma(S\to \gamma\gamma)\Gamma(S\to gg)}{\Gamma_{S;\text{total}}}
\nonumber\\
&\simeq
6.4~\text{fb}\times \lrf{\Gamma(S\to \gamma\gamma)}{1~\MeV}\text{Br}(S\to gg)
\end{align}
where $m_S\simeq 750~\GeV$ is the scalar mass, 
$\sqrt{s}=13\ {\rm TeV}$ and $C_{gg}=(\pi^2/8)\int^1_0 dx_1 \int^1_0 dx_2$
$\delta(x_1x_2-m_S^2/s)g(x_1)g(x_2)$ with $g(x)$ being the gluon
parton distribution function.  In the second equation, we have used $C_{gg}\simeq 2.1\times 10^3$
which is obtained from MSTW2008 NLO set~\cite{Martin:2009iq} evaluated at the
scale $\mu=m_S$. The reported signal strengths at Run 2 are somewhere between 1  and 10 fb \cite{ATLAS:diphoton,CMS:2015dxe}.

We assume that  couplings of the scalar $S$ to photons and gluons are induced by 
 a VLQ $X$ which transforms as $({\bf 3}, {\bf 2}, 7/6)$ under 
${\rm SU}(3)_{\rm c}\times {\rm SU}(2)_{\rm L}\times {\rm U}(1)_{\rm Y}$ of the SM gauge group:
\begin{align}
-{\cal L}_{\text{int}} = y S\bar{X} i\gamma_5 X\,,
\end{align}
where $S$ is assumed to be CP-odd.\footnote{In the case of CP-even $S$, the diphoton signal rate is suppressed by a factor of about $4/9$.}
Then, the partial decay width of $S$ into the diphoton is given by
\begin{align}
\Gamma(S\to \gamma\gamma)
&=
\frac{\alpha^2}{64\pi^3}
m_S^3
\left[
 \frac{y}{m_X} \text{Tr}( Q^2)\ 
 f(m_S^2/4m_X^2)
\right]^2 
\nonumber\\
&\simeq 
1.4~\MeV\times y^2 \lrfp{1~\TeV}{m_X}{2}  f(m_S^2/4m_X^2)
\end{align}
where $\text{Tr}(Q^2)=(5/3)^2+(2/3)^2=29/3$ and the loop function is given by 
$f(\tau)=\tau^{-1} \arcsin^2\sqrt{\tau}$~\cite{Djouadi:2005gj}.

%%%%%%%%%%%%%%%%%%%%%%%%%%%%%%%%%%%%%%%%%%%
\begin{figure}[t]
\includegraphics[width=0.45\textwidth]{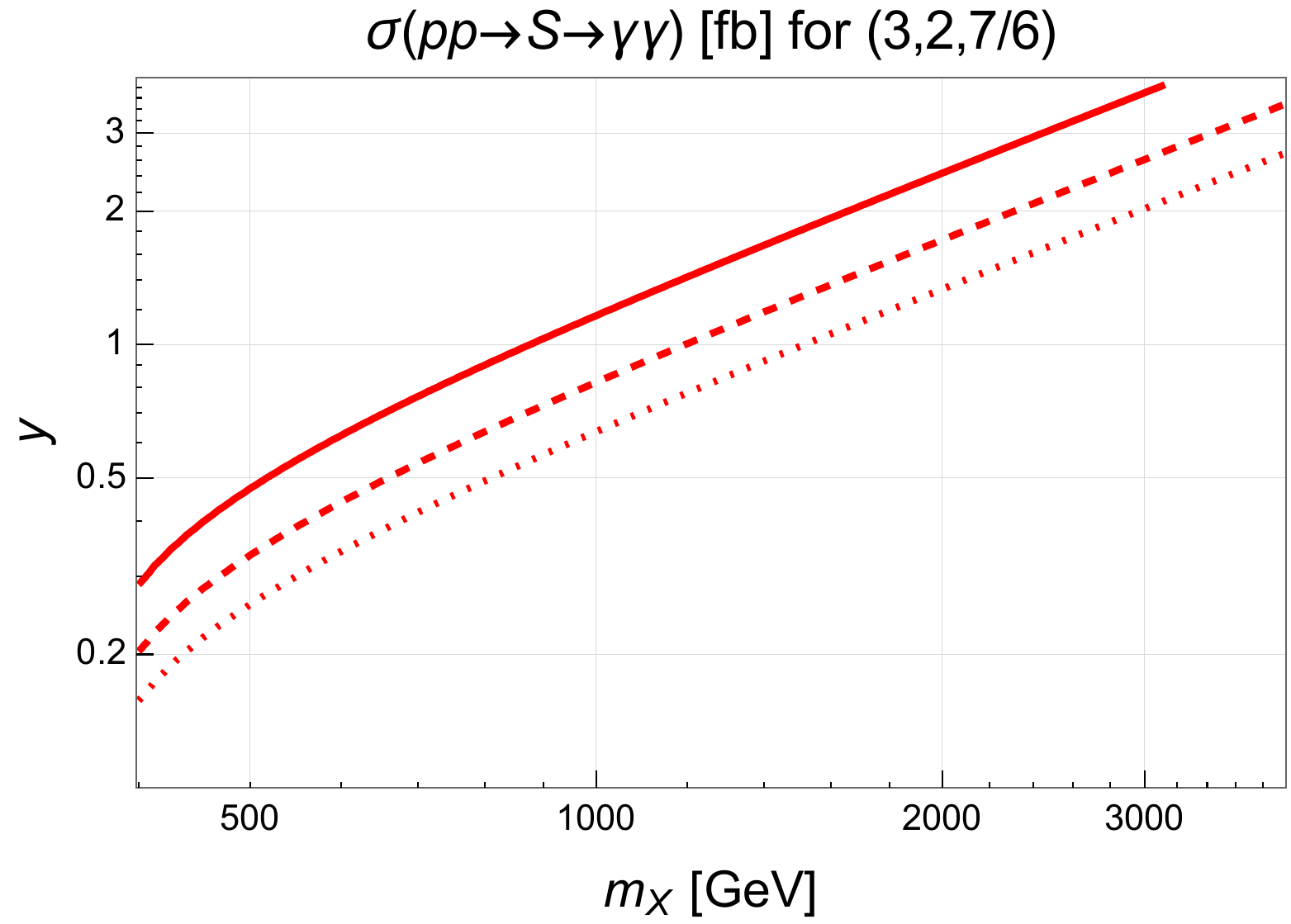}
\includegraphics[width=0.45\textwidth]{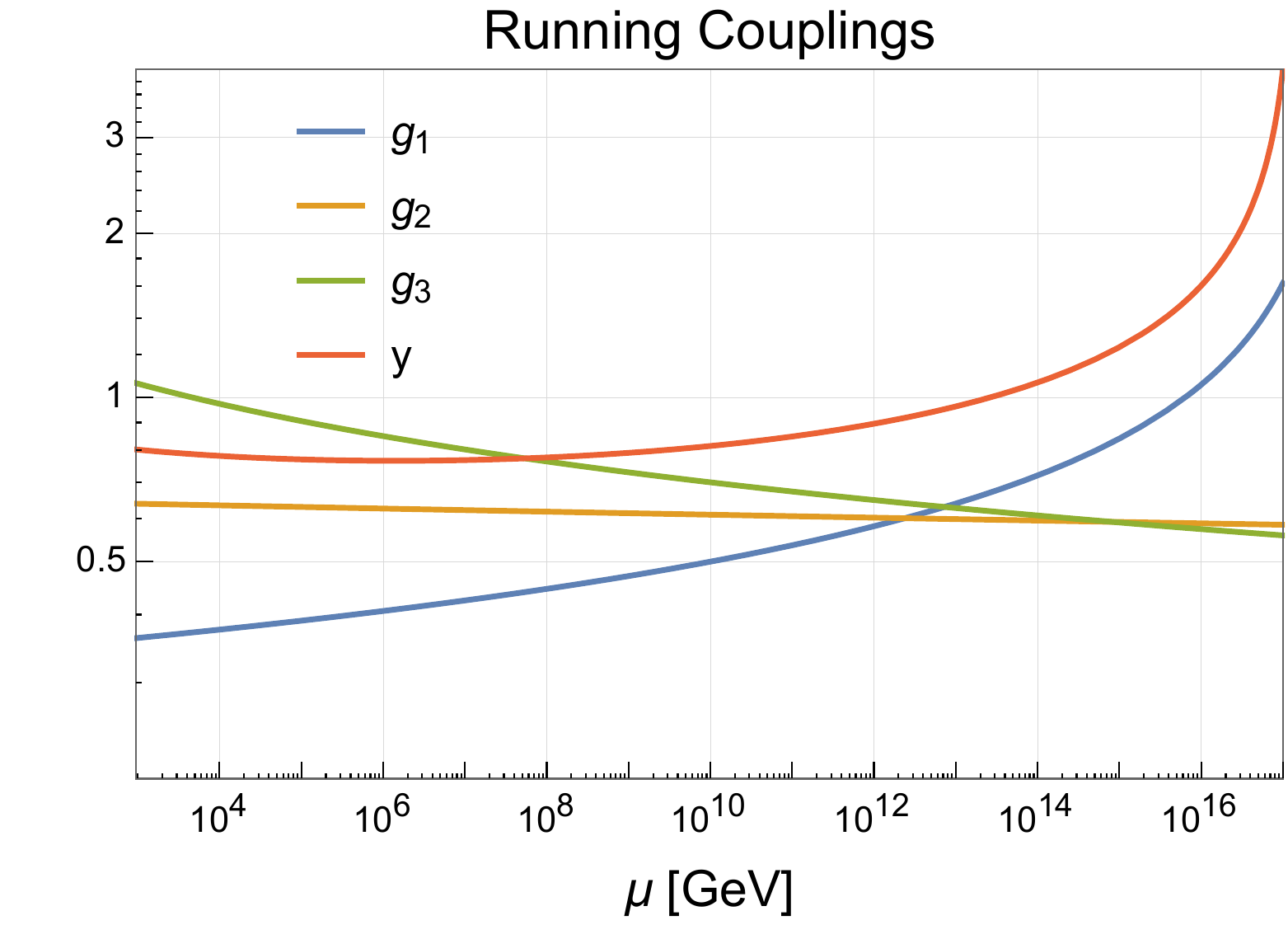}
\caption{\em Left: contours of 
$\sigma(pp\to S\to \gamma\gamma)$ in $(m_X, y)$ plane. 
Solid, dashed and dotted lines respectively show contours for 10, 5, 3 fb.
Right: the running of the SM gauge couplings and the Yukawa coupling $y$
for a benchmark point $m_X=950~\GeV$ and $y=0.8$.
 }{\label{fig:diphoton_and_running}}
\end{figure}
%%%%%%%%%%%%%%%%%%%%%%%%%%%%%%%%%%%%%%%%%%%

In Fig.~\ref{fig:diphoton_and_running}, we show contours of 
$\sigma(pp\to S\to \gamma\gamma)$ in $(m_X, y)$ plane.
The observed diphoton signal rate can be obtained with $m_X\alt 1~\TeV$
and $y\alt {\cal O}(1)$. In the right panel of Fig.~\ref{fig:diphoton_and_running},
the running of the SM gauge couplings and the Yukawa coupling $y$ are shown 
for a benchmark point $m_X=950~\GeV$ and $y=0.8$, which leads to
$\sigma(pp\to S\to \gamma\gamma)\simeq 5.4$~fb.
As can be seen in the figure, the model remain perturbative up to about $10^{17}~\GeV$.\footnote{If we add an additional VLQ $T'$, which is introduced in the model 3 discussed in the next section, the Landau pole of the U(1)$_Y$ coupling becomes slightly lower, about $10^{16}~\GeV$.}

%%%%%%%%%%%%%%%%%%
\section{Vector-Like Quarks and Multilepton Excess}
\label{sec:VLQ_and_Multilepton}

The VLQ $X$, introduced in Sec.~\ref{sec:VLQ_and_diphoton} to explain the diphoton excess, 
cannot be stable nor long-lived in order to avoid severe collider bound~\cite{Aaboud:2016dgf}.  We will assume  the lifetime of the VLQ is short enough to decay promptly inside the detector,  when produced through QCD interactions at the LHC. Given the $X$ carries SM quantum numbers, its decay product would include SM particles such as quarks and gauge bosons. In particular, if the charge-5/3 VLQ, $X_{5/3}$, decays into the top quark and the $W$ boson, it could potentially contribute to the multilepton excesses observed at both Run 1 and Run 2 of the LHC.

We consider two simplified models in the spirit of minimality, one with a dark matter candidate and one without. Matter contents of both models are shown in Table.~\ref{table:matter_contents}. In the first model,  Model S, the only new particles are the $S$ and $X$ which are introduced in Sec.~\ref{sec:VLQ_and_diphoton}. We assume that the $X$ field mainly couples to the third generation SM quark:
\begin{align}
-{\cal L}_{\text{int}}^{\text{Model S}}
&= -\lambda (H\cdot \overline{X}) t_R + h.c. + y S\bar{X} i\gamma_5 X\,.
\end{align}
where $t_R$ is the right-handed SM top quark and $(H\cdot \bar{X})=H^+ \overline{X}_{5/3} + H^0 \bar{T}$
with $(H^+,H^0)$ and $(X_{5/3},T)$ being the component fields of the SU(2)$_{\rm L}$ doublets.
In this setup, the decay chain of the VLQ $X_{5/3}$ is given by
\begin{align}
X_{5/3}\to W^+ t \to W^+ W^+ b\,.
\end{align}
Therefore, pair-produced $X_{5/3}$ can lead to signatures containing multilepton, $b$-jets, and MET, where the MET comes from the neutrino in the $W$ decay. 

%%%%%%%%%%%%%%%%%%
\begin{table}[t]
\begin{tabular}{|ll|lll|}
\hline
\multicolumn{2}{|l|}{Model S}
&
\multicolumn{3}{|l|}{Model 3}
\\ \hline
$S$ & $({\bf 1}, {\bf  1})_0$ &
$S$ & $({\bf 1}, {\bf  1})_0$ & $+$
\\
$X$ & $({\bf 3},{\bf 2})_{7/6}$ &
$X$ & $({\bf 3},{\bf 2})_{7/6}$ & $-$
\\
&&
$T'$ & $({\bf 3},{\bf 1})_{2/3}$ & $-$
\\
&&
$\phi$ & $({\bf 1}, {\bf  1})_0$ & $-$
\\ \hline
\end{tabular}
\caption{\em Matter contents of two simplified models. Model 3 contains a new parity under which the SM and $S$ are even, while $X=(X_{5/3},T)$, $T'$ and $\phi$ are odd. $\phi$ is the lightest particle charged under the parity and could serve as a dark matter candidate $\phi$.}
\label{table:matter_contents}
\end{table}
%%%%%%%%%%%%%%%%%%

\begin{figure}[t]
\includegraphics[width=0.4\textwidth]{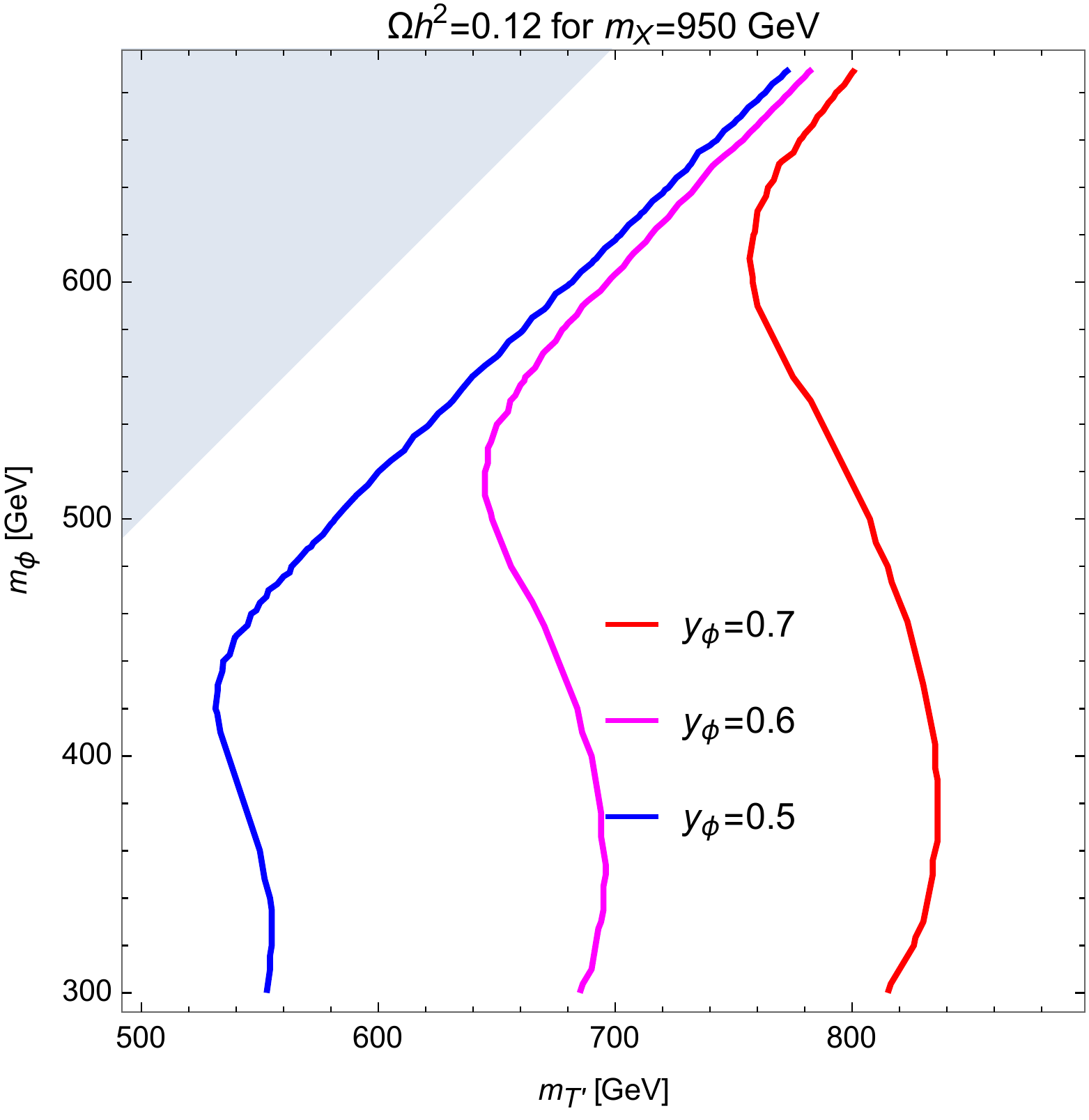}
\caption{\em Region of parameter space of Model 3 leading to the observed dark matter relic density.
 }{\label{fig:dmrelic}}
\end{figure}

In the second model containing the dark matter, Model 3, we introduce a singlet scalar dark matter field $\phi$ together with an additional VLQ, $T'$, and  assign an odd parity to $X=(X_{5/3},T)$, $T'$, and $\phi$ under a $Z_2$ parity to ensure the stability of the dark matter. The interaction is then given by\footnote{The $T'$ field can also have a coupling with $S$ as $y' S\bar{T'} i\gamma_5 T'$, which slightly increases the diphoton rate.}
\begin{align}
\label{eq:model3}
-{\cal L}_{\text{int}}^{\text{Model 3}}
&=  (H\cdot \overline{X})(\lambda + \lambda_5 \gamma^5) T' + y_\phi\ \phi \bar{T'} t_R + h.c.
+ y S\bar{X} i\gamma_5 X\,.
\end{align}
In the following, we assume $\lambda_5=0$, for simplicity.
In this model the decay of $X_{5/3}$ leads to
\begin{align}
X_{5/3}\to W^+ T' \to W^+ t \phi \to W^+ W^+ b \phi\,,
\end{align}
which again leads to the multilepton signal, with the additional MET contribution from the DM candidate $\phi$. Furthermore, $T'$ decays as follows
\be
T'\to t+\phi \ ,
\ee
therefore pair-production of $T'$ at the LHC could lead to the final state $t\bar{t} + {\rm MET}$. In the early universe, the dark matter $\phi$ could annihilate through the process $\phi\phi\to t\bar t$. In Fig.~\ref{fig:dmrelic} we show the region of parameter space leading to the observed relic density of the dark matter in the $m_\phi - m_{T'}$ plane for some choices of $y_\phi$.\footnote{The $\lambda$ term in Eq.~(\ref{eq:model3}) induces a mixing between $T'$ and $T$ after electroweak symmetry breaking. We choose a small mixing, $\lambda=0.01$ in Fig.~\ref{fig:dmrelic}.}  For numerical calculation, we generate model files using {\tt FeynRules~2.3}~\cite{Alloul:2013bka} and obtain the relic density of dark matter using {\tt micrOMEGAs~4.1.7}~\cite{Belanger:2014vza}. In addition, in this simple model the only SM particle that directly couples to the dark matter particle $\phi$ is the top quark. Therefore the direct detection constraint does not apply.

%%%%%%%%%%%%%%%%%%%%%%%%%%%%%%%%%%%%%%%%%%%
%%%%%%%%%%%%%%%%%%%%%%%%%%%%%%%%%%%%%%%%%%%

There are collider searches for $X_{5/3}\to tW^+$ at the LHC. In this channel the latest result from CMS using the early Run 2 data puts a lower bound $m_{X}\agt 950$ GeV at the 95\% confidence level (C.L.)~\cite{CMS:2016rnk}, which would apply to the $X_{5/3}$ in Model S. In Model 3 $X_{5/3}$ decays into $tW^+\phi$ and the strong bound from CMS does not necessarily apply since the decay final state is different. The charge $2/3$ component of the doublet, $T$, decays mainly to $t + Z/h$, and the current constraint is weaker than that of $X_{5/3}$~\cite{Aad:2015mba,Khachatryan:2015oba}.  For the electroweak singlet $T'$, no dedicated searches exist, although there exist searches for the scalar top quark (stop) in supersymmetry which also decays to $t\bar{t}+{\rm MET}$. These limits, however, are sensitive to a variety of kinematic variables including the mass of the particle carrying away the extra MET. In particular, constraints on direct stop productions in the $t\bar{t}$+MET final states disappear when the LSP mass becomes larger than $\sim 300$ GeV \cite{Aad:2015pfx,ATLASstop,Khachatryan:2016pup,CMS:2016qtx}. Due to these considerations, we choose the following benchmark in our study
\begin{align}
&\text{Model S:}\quad m_X = 950 {\rm \ GeV} \ ,
\\
&\text{Model 3:}\quad m_X = 950 {\rm \ GeV} \ ,\quad m_{T'} = 750 {\rm \ GeV}\ , \quad m_{\phi}= 380 {\rm \ GeV} \ .
\end{align}
The choice of $m_X$ in Model S maximizes the contribution to the multilepton excess without contradicting the limits from searches for $X_{5/3}$ at the LHC,\footnote{We emphasize that the CMS bound on $X_{5/3}$ from Ref.~\cite{CMS:2016rnk} does not apply to our Model 3. However, we choose the same $m_X$ in both models for simplicity.} while the value for $m_\phi$ is motivated not only by the stop constraints but also the need to turn off possible tree-level decay $S\to \phi\phi$,\footnote{Such a decay could arise from the interaction $S\phi\phi$, which is irrelevant for our study for $m_\phi\ge m_S/2$.} so as not to overwhelm the loop-induced decays into diphotons. $m_{T'}$ is then chosen to allow for on-shell $W$ bosons and top quarks in the decay product.

In Table \ref{tab:signalstrength} we show the total signal strength, in unit of SM expectations, of our two benchmarks in the ttH and ttW channels at the LHC. In the Monte Carlo simulation we generate signal events using {\tt Madgraph} \cite{Alwall:2007st} + {\tt Pythia} \cite{Sjostrand:2006za} + {\tt Delphes} \cite{deFavereau:2013fsa}  and a tagging efficiency of $b$-jet at 77\%. In the ttH channel we implement the same cuts as in Ref.~\cite{Chen:2015jmn} at both 8 TeV and 13 TeV, and normalize the resulting events to the number expected from the SM ttH production. In the ttW channel we base our selection cuts on those in the Section 5.1 of Ref.~\cite{atlasrun2ttw} in the same-sign di-muon channel. The total signal strength is then the sum of the SM expectation and the contribution from the $X_{5/3}$ VLQ,
\be
\mu \equiv \frac{\sigma_{\rm SM}+\sigma_{\rm VLQ}}{\sigma_{\rm SM}} = 1 + \mu_{\rm VLQ}  \ ,
\ee 
which is shown in Table  \ref{tab:signalstrength}. For comparison, we note that the 95\% C.L. upper limit in the ttH multilepton channel from CMS Run 2 data is \cite{CMS:2016rnk}
\be
\mu_{\rm ttH} < 3.3 \ (2.6\ {\rm expected}) \quad \text{at 95\% C.~L.}
\ee
We see that both benchmarks give good fits to the multilepton excesses observed at both Run 1 and Run 2.

%%%%%%%%%%%%%%%%%%%%%%%%%%%%%%%%%%%%%%%%%%%
\begin{table}[t]
\begin{tabular}{|c||c|c|c|}\hline
 & 8 TeV $\mu_{\rm ttH}$  & 13 TeV $\mu_{\rm ttH}$ & 13 TeV $\mu_{\rm ttW}$ \\
 \hline\hline
 Model S & 1.5  & 3.1 & 1.3 \\
 \hline
Model 3 & 1.4   & 2.9 & 1.3 \\
\hline
\end{tabular}
\caption{\em Total signal strengths, normalized to the SM expectations, of our benchmark models in {\rm ttH} channel at both 8 TeV and 13 TeV and in {\rm ttW} channel at 13 TeV. The mass spectrum is $m_X=950$ GeV, $m_{T'}$=750 GeV and $m_{\phi}$=380 GeV.}
\label{tab:signalstrength}
\end{table}
%%%%%%%%%%%%%%%%%%%%%%%%%%%%%%%%%%%%%%%%%%%

\begin{figure}[t]
\includegraphics[width=0.32\textwidth]{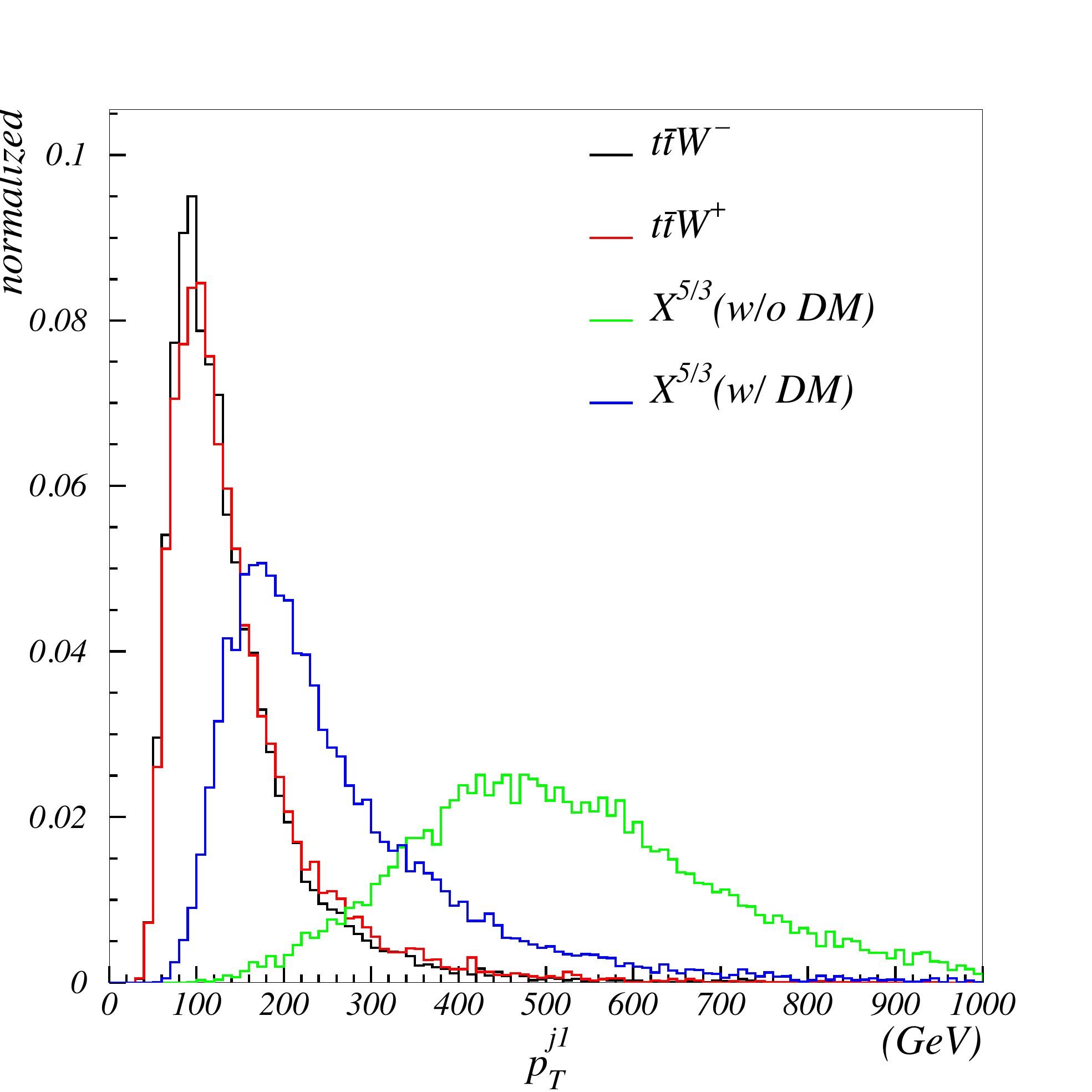}
\includegraphics[width=0.32\textwidth]{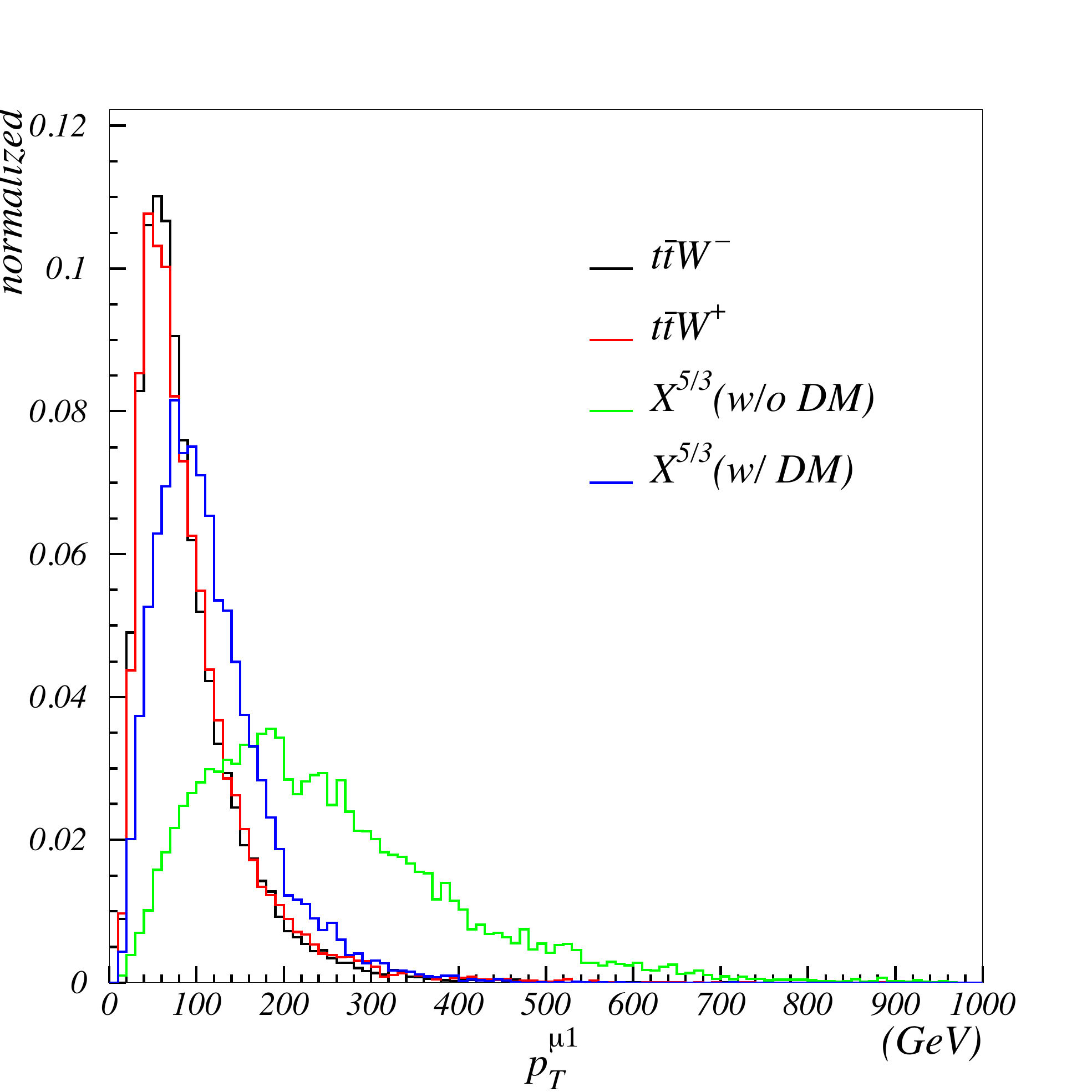}
\includegraphics[width=0.32\textwidth]{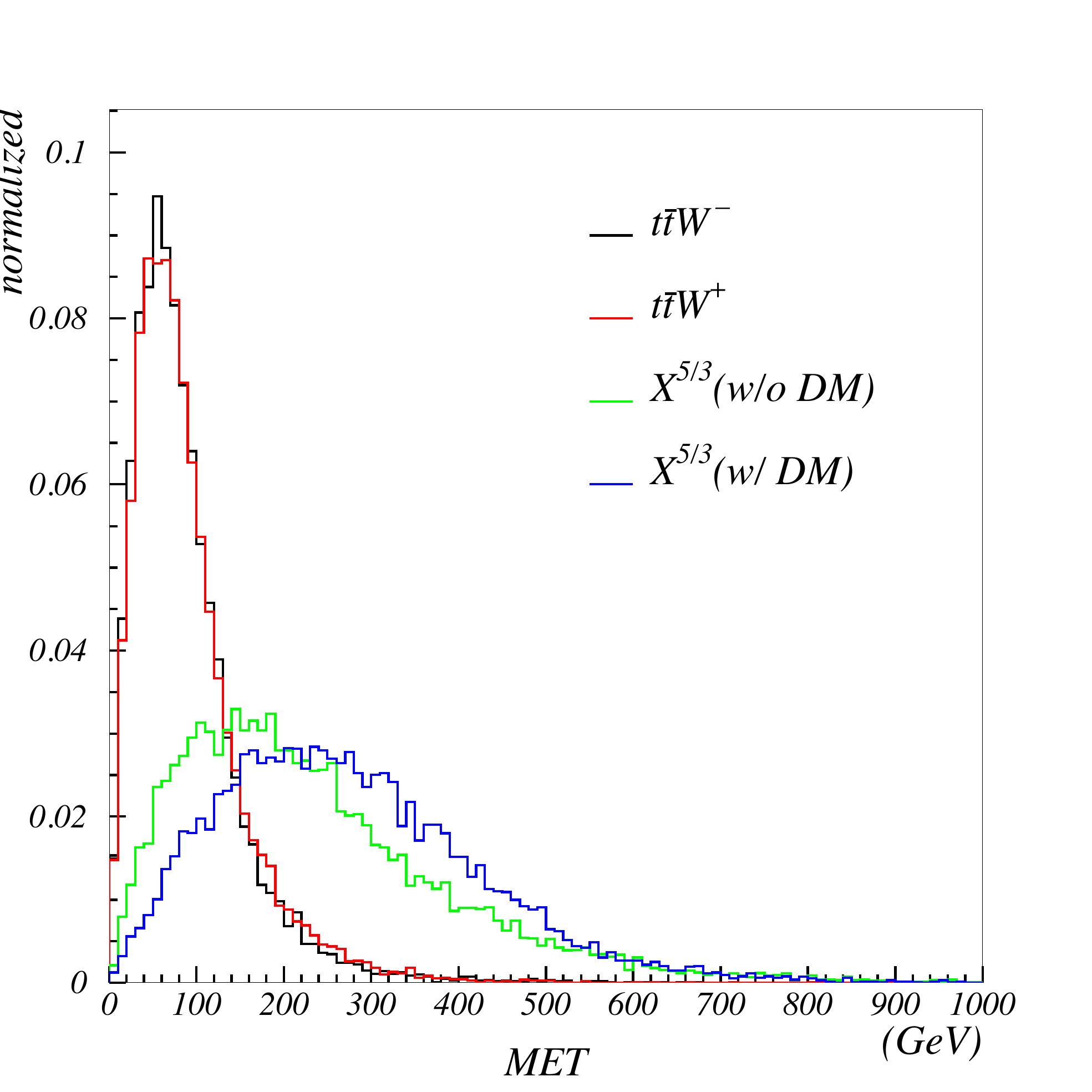}
\caption{\em From left to right: distributions of $p_T$ of the leading jet, $p_T$ of the leading muon, and MET for events from our Model S ($X_{5/3}$ without the dark matter), Model 3 ($X_{5/3}$ with the dark matter), and SM ttW. 
 }{\label{fig:kindist}}
\end{figure}

In the future, should the multilepton excess persist, it would be crucial to understand the nature of the excess. In this regard, we compare some kinematic distributions of events in the same-sign di-muon channel from our benchmark models with those from SM ttW in Fig.~\ref{fig:kindist}. In particular, we plot $p_T$ of the leading jet and the leading muon, respectively, as well as the MET distribution. It is clear that events  from our benchmarks  in general have harder decay spectra than those from the SM. 

Fig.~\ref{fig:kindist} also highlights the importance of conducting dedicated searches for VLQs whose decay product contains a dark matter candidate carrying away  extra MET, as the kinematic distributions are quite different between the two benchmarks. Such searches have not been performed at the LHC to the best of our knowledge. Although the MET distribution for Model 3 is harder than Model S, $p_T$ distributions of the leading jet and the leading muon are softer in Model 3. Therefore we expect the selection efficiency will be different between the two benchmarks.

%%%%%%%%%%%%%%%%%%%%%%%%%%%%%%%%%%%%%%%%%%%
%%%%%%%%%%%%%%%%%%%%%%%%%%%%%%%%%%%%%%%%%%%

%%%%%%%%%%%%%%%%%%
\section{Conclusion}
%%%%%%%%%%%%%%%%%%%%%%%%%%%%%%%%%%%%%%%%%%%

In this work we studied the LHC phenomenology of the VLQ that is often invoked to mediate loop-induced couplings of the 750 GeV diphoton resonance. In particular, we considered an electroweak doublet VLQ with hypercharge 7/6, giving rise to an exotic charge-5/3 VLQ and a top-like VLQ with charge-2/3. When the mass of the VLQ is at around 1 TeV, the desired diphoton signal strength can be obtained for a Yukawa coupling $y\alt {\cal O}(1)$. In this scenario, all gauge couplings as well as the VLQ Yukawa coupling remain perturbative up to a very high energy scale, of the order of $10^{17}$ GeV.

Decay phenomenology of the VLQ is quite distinct at the LHC, especially if they decay into the third generation fermions and the SM gauge bosons. For example, the charge-5/3 VLQ, $X_{5/3}$, could decay into a top quark and a $W^+$ boson. Alternatively, it is possible to include a stable neutral particle as the dark matter candidate with the correct relic density. Then the VLQ always decays into SM particles and the dark matter particle, which carries away additional MET in the collider detectors. Interestingly, we demonstrated that decays of $X_{5/3}$ could contribute to the mild excess in the multilepton, $b$-jets, and MET channel that are observed at both Run 1 and Run 2 of the LHC.

We performed numerical studies on two benchmark models, one with the dark matter particle and one without, and compared their kinematic distributions with those from the SM ttW processes. Should the multilepton excess persists in the future, such comparisons  will shed light on the nature of the excess. Furthermore, we showed that kinematic distributions between the two benchmarks are  quite distinct, calling out the need  for dedicated experimental efforts to search for a new decay topology of VLQs,  into final states containing a dark matter candidate.

%%%%%%%%%%%%%%%%%%
\section*{Acknowledgement} 
The authors are grateful to the workshop
``Beyond the Standard Model in Okinawa 2016," March 1 - 8, 2016, OIST, Okinawa, Japan, 
where this work was initiated. 
I. L. acknowledges helpful discussions with Bill Murray and the hospitality of  KITP in Santa Barbara, which is supported by the National Science Foundation under Grant No. NSF PHY11-25915.
This work was supported in part by Grant-in-Aid for Scientific research 
(Nos.\ 26104001, 26104009, 26247038, 26800123, 16H02189), World Premier International
Research Center Initiative (WPI Initiative), MEXT, Japan, the U.~S. Department of Energy under contracts No. DE-AC02-06CH11357 and No. DE-SC 0010143, and the National Science Council of R.O.C. under Grants No. NSC 102-2112-M-003-001-MY3. 
%%%%%%%%%%%%%%%%%%%%%%%%%

\bibliographystyle{utphys}
\bibliography{diphoton_ss2l}

\end{document}